\documentclass[acmsmall,screen,authorversion,nonacm]{acmart}

\AtBeginDocument{%
  \providecommand\BibTeX{{%
    \normalfont B\kern-0.5em{\scshape i\kern-0.25em b}\kern-0.8em\TeX}}}

\setcopyright{acmcopyright}
\copyrightyear{2021}
\acmYear{2021}
\acmDOI{0000001.0000001}

\acmJournal{TIIS}
\acmVolume{00}
\acmNumber{0}
\acmArticle{000}
\acmMonth{0}

\usepackage{_macros}
\usepackage{caption}
\usepackage{subcaption}
\hypersetup{draft}

\newcommand{\name}{EDAssistant}

\begin{document}

\title[\name]{\name: Supporting Exploratory Data Analysis in Computational Notebooks with In-Situ Code Search and Recommendation}


\author{Xingjun Li} \authornote{These authors contributed equally to this work.}
\author{Yizhi Zhang} \authornotemark[1]
\author{Justin Leung} \authornotemark[1]
\author{Chengnian Sun}
\author{Jian Zhao}
\affiliation{%
  \institution{University of Waterloo}
  \city{Waterloo}
  \state{Ontario}
  \country{Canada}}
\email{{xingjun.li,yizhi.zhang,justin.leung1,cnsun,jianzhao}@uwaterloo.ca}

\renewcommand{\shortauthors}{Li, et al.}

\begin{abstract}
Using computational notebooks (\eg, Jupyter Notebook), data scientists rationalize their exploratory data analysis (EDA) based on their prior experience and external knowledge such as online examples.  
For novices or data scientists who lack specific knowledge about the dataset or problem to investigate, effectively obtaining and understanding the external information is critical to carry out EDA. 
This paper presents \name{}, a JupyterLab extension that supports EDA with in-situ search of example notebooks and recommendation of useful APIs, powered by novel interactive visualization of search results.  
The code search and recommendation are enabled by state-of-the-art machine learning models, trained on a large corpus of EDA notebooks collected online.
A user study is conducted to investigate both \name{} and data scientists' current practice (\ie, using external search engines). 
The results demonstrate the effectiveness and usefulness of \name{}, and participants appreciated its smooth and in-context support of EDA. We also report several design implications regarding code recommendation tools. 
\end{abstract}



\maketitle

\section{Introduction}

Exploratory data analysis (EDA) \cite{tukey1977exploratory} is a critical process in modern data science workflows \cite{batch2018interactivea, barel2020automatically, subramanian2020tractus}. 
During EDA, data scientists investigate new datasets or problems with the broader goal of understanding ``what is going on here?'' and with an emphasis on visualization of data, iterative and tentative model building, as well as hypothesis generation and measures \cite{behrens1997principles}.
Due to its vague goal and exploratory nature, EDA is often challenging, as data scientists need to decide among a large number of possible actions \cite{battle2019characterizing,wongsuphasawat2017voyager, zhao2020chartseer}.
To mitigate this challenge, many visual tools have been proposed to facilitate the EDA process with intuitive authoring interfaces (\eg, Tableau\footnote{https://www.tableau.com/} and PowerBI\footnote{https://powerbi.microsoft.com/}), data wrangling support \cite{kandel2011wrangler}, and recommendations of data visualizations \cite{wongsuphasawat2017voyager,zhao2020chartseer}.  

Despite the benefits of these visual tools, computational notebooks, such as Jupyter Notebook\footnote{https://jupyter.org/} and RStudio\footnote{https://www.rstudio.com/}, are the single most popular and useful means for data scientists to perform EDA \cite{rule2018exploration,kery2018story}.
One reason is that computational notebooks combine code, documentation, and outputs (\eg, tables, charts, and images) within a single document, which provides expressive and interactive support for EDA.
Moreover, computational notebooks support literate programming in languages such as Python, which allows for directly integrating EDA code into a production pipeline. 
Most importantly, computational notebooks can be easily shared and hosted on platforms such as GitHub\footnote{https://github.com/} and Kaggle\footnote{https://www.kaggle.com/}, making it possible to leverage the collective wisdom of the data science community. 

Indeed, people learn programming from the widely-available examples and tutorials online, and code search is one frequent and critical activity for developers \cite{brandt2009two,robillard2010field,sim2011how,zhang2018are}.   
This is the same for EDA and using notebooks \cite{head2020composing,subramanian2020tractus}. 
When data scientists, especially novices, start to investigate their data or are stuck at some problems, they usually look for EDA notebooks online to learn how others approach the same or similar problem, check the APIs, models, and metrics that others have used, and get inspirations for performing EDA themselves \cite{rule2018exploration,subramanian2020tractus}.
The code that data scientists write in their current notebooks is often an externalization of their thoughts or hypotheses, which could inform future steps.  
Thus, leveraging large repositories of EDA notebooks, several researchers have attempted to use machine learning for automating EDA, such as recommending next operations in data wrangling \cite{yan2020autosuggest} and generating EDA sessions from a dataset \cite{barel2020automatically}. 
However, these methods are constrained to a small set of EDA operators (\eg, \texttt{filter}, \texttt{merge}, and \texttt{groupby} in pandas\footnote{https://pandas.pydata.org/}), sometimes imprecise due to the complexity of EDA goals, and lack human engagement and interaction.  
Further, while these methods sometimes could inform data scientists what to do next, they do not tell the data scientists why or allow them to learn or improve skills.

To fill the gap, we propose \name{}, an interactive and visual tool that facilitates EDA with in-situ code search, exploration, and recommendation based on existing notebook repositories, embedded within the JupyterLab environment for a seamless user experience.  
To design \name{}, we first curated a large corpus of EDA notebooks (consisting of 38,581 notebooks from Kaggle), and characterized the EDA process based on a formative study with two data scientists as well as a quantitative analysis of the corpus. 
We confirmed the observation that data scientists organize their EDA code in sequences of blocks (\eg, \cite{subramanian2020tractus,li2021nbsearch}), and discovered four major types of EDA blocks. 
Our findings also reflect typical characteristics of data science workflows mentioned in the literature \cite{batch2018interactivea,kandel2012enterprise}.
With the analysis and dataset, we employed advanced deep learning models, specially GraphCodeBERT \cite{guo2020graphcodebert}, to learn a latent representation (\ie, embeddings) of all the EDA sequences. 
As the backend of \name{}, we developed a search engine for retrieving relevant EDA sequences based on a data scientist's current code and a recommender for potential APIs to use next, which facilitates their EDA with useful examples and suggestions.  
Further, we built a visual interface, as a JupyterLab extension, to allow users to conduct EDA while accessing \name{} smoothly.  
The user interface also features a novel visualization that provides an informative overview of the search results and the coding patterns in EDA notebooks. 

During the development of \name{}, we conducted quantitative experiments to compare different models including GraphCodeBERT and Doc2Vec \cite{le2014distributed} in our search and recommendation tasks. 
We also carried out a user study with 14 participants, who have different levels of technical expertise in data science, to evaluate \name{} as a whole on conducting EDA, by referencing to a baseline setting of using external search engines (\eg, Google). 
The results indicate that participants appreciated the new in-situ code search and recommendation experience during EDA as well as the interface design of \name{} on retrieving, exploring, and understanding code examples. 
Moreover, several design implications are discussed, shedding light on future research. 
For example, participants sometimes benefited from the diverse results on Google (\eg, video tutorials, forum discussions, etc.), and combining \name{}'s in-situ search with traditional manual keyword-based search could potentially improve their EDA performance. 

In summary, our contributions in this paper include:
1) an empirical characterization of the EDA process with a formative study and a quantitative analysis of a large notebook corpus;
2) a search engine for retrieving EDA examples and a recommender for suggesting useful APIs based on the application of the state-of-the-art machine learning models;
3) an interactive tool, implemented within the Jupyter Notebook environment, which offers in-situ code search and recommendation as well as novel visual exploration of search results\footnote{We will make the corpus and code publicly available upon publication.}.


\section{Background}
In this section, we first introduce computational notebooks and relevant tools, then exploratory data analysis and systems, and lastly techniques for searching, recommending, and visualizing code in general.

\subsection{Computational Notebooks}
Computational notebooks (\eg, Jupyter Notebook and RStudio) have recently emerged as a new form of programming environment. 
A notebook is broken down into \textit{cells}, which contains \textit{code cells} that are segments of scripts, and \textit{markdown cells} that are formatted text to supplement and explain the code. 
Code cells can be independently executed in an arbitrary order, run multiple times, or even edited between different runs; these cells are also interrelated as in the same notebook environment where they sharing common variables, function definitions, and so forth. 
In addition, the outputs of executed code cells, such as charts, tables and printouts, are embedded in the notebook in place. 
The above characteristics of notebooks provide much freedom and flexibility, which perfectly suit EDA, allowing data scientists to dynamically experiment with different methods, try out alternative models, and write temporary code~\cite{kery2018story,rule2018exploration}.   
However, at the same time, such flexibility creates challenges in code management, comprehension, and development with notebooks. 
For example, messes in code may accrue during EDA, and data scientists may lose track of their thought processes. 
To address these issues, several tools have been proposed, such as Variolite \cite{kery2017variolite}, Verdant \cite{kery2019effective}, and ForkIt \cite{weinman2021fork}, to support fast versioning and history tracking.
Code Gathering Tools \cite{head2019managing} assist data scientists with cleaning, recovering, and comparing versions of code in notebooks by analyzing code cells' dependency and organizing code into chunks.

However, these tools focus on general code management and versioning in notebooks, rather than EDA tasks. 
In our work, by taking the huge advantages of computational notebooks in supporting EDA, we designed and developed \name{} as a JupyterLab extension, which can be used seamlessly in the environment that is familiar to data scientists.  
Further, \name{} facilitates EDA in notebooks with in-situ code search, exploration, and recommendation, enabled by analyzing a large collection of EDA notebooks gathered online.  

\subsection{Exploratory Data Analysis and Tools}
The concept of exploratory data analysis (EDA) stems from John Tukey's early work \cite{tukey1977exploratory}. 
The nature of EDA is loosely characterized by addressing goals or hypotheses with skepticism and flexibility, emphasizing the use of data visualization, building tentative models, and applying robust measures, in an iterative manner \cite{behrens1997principles}. 
Batch \etal~\cite{batch2018interactivea} further advocated the gap in using interactive visualization for EDA in data science. 

Data scientists need to make many decisions during EDA, such as which models to employ, which parts of data to examine, and which graphic representations to use? Thus, a number of techniques have been proposed to facilitate EDA, where most focus on supporting the creation of data visualization. 
Commercial tools like Tableau and PowerBI allow analysts to interactively explore data, further enabled by intelligent algorithms (\eg, Tableau ``Show Me'' \cite{mackinlay2007show}) to recommend expressive visualization. 
Visualization specification languages, such as Vega-Lite \cite{satyanarayan2017vega}, have also been proposed to ease the process of creating common data charts programmatically.  
Based on Vega-Lite, Voyager \cite{wongsuphasawat2016voyager} and Voyager 2 \cite{wongsuphasawat2017voyager} blend manual and automated visualization specifications in EDA. 
Falx \cite{wang2021falx} automatically infers visualization specification and data transformation from user-input examples. 
Another branch of research uses data-driven methods for automatically generating visualization. 
Examples include Data2Vis \cite{dibia2019data2vis}, DeepEye \cite{luo2018deepeye}, and VisML \cite{hu2019vizml}, which employ deep learning techniques to extract rules, patterns, and designs from large collections of user-created charts.
Further, ChartSeer \cite{zhao2020chartseer} adopts a mixed-initiative approach to recommend visualization designs dynamically based on an analyst's input. 
While the above systems are effective in creating visualization, they are standalone tools separated from the computational notebook environment.
Further, EDA is more than just data visualization \cite{behrens1997principles}, which also includes data processing, model building and evaluation, etc. Thus, switching among tools for different EDA tasks significantly reduces the effectiveness of a data scientist's workflow. 

There thus exist several tools, created with friendly integration to computational notebooks, for supporting different aspects of EDA.
Many R and Python packages have been developed, such as tidyverse\footnote{https://www.tidyverse.org/} (containing ggplot2, etc.), matplotlib\footnote{https://matplotlib.org/}, scikit-learn\footnote{https://scikit-learn.org/stable/}, and pandas, which can be directly used in the corresponding notebook environments.
There also exist different techniques proposed as computational notebook extensions or plugins. 
BURRITO \cite{guo2012burrito} and TRACTUS \cite{subramanian2020tractus} provide the provenance of EDA by capturing and displaying code outputs, development activities, and users' hypotheses.
Wrex \cite{drosos2020wrex} is a Jupyter Notebook extension that supports data wrangling with a programming-by-example approach. 
B2 \cite{wu2020b2} allows users to easily move from code to visualization and vice versa by treating data queries as a shared representation.
Although providing much convenience, these tools assume that data scientists have a good idea about what to do in their EDA, lacking the support of ``tutoring or inspiring'' them, especially for less-experienced data scientists.
Often they still need to leverage other means, such as Google Search, to find, browse, and learn from example notebooks online, which is our focus in this work. 

Learning from a large collection of notebooks, Auto-Suggest \cite{yan2020autosuggest} can recommend the next step (\eg, which pandas API to use, and with what parameters), but only in the data processing phase of EDA. 
Similarly, ATENA \cite{barel2020automatically} automatically generates data exploration sessions using deep reinforcement learning; however, the types of EDA operations are limited to data filtering and grouping.
Inspired by these data-driven methods, \name{} leverages the collective wisdom of the data science community by analyzing high-quality EDA notebooks online to facilitate EDA with in-place code search and API recommendation. 
Instead of viewing EDA as a series of low-level operations, we treat EDA as a sequence of semantic code blocks to enhance the utility of the search results.
Also, different from these pure automated methods, we employ interactive visualization to allow data scientists to explore and understand searched notebooks, thus better utilizing the examples and gaining knowledge.

\subsection{Code Search and Visualization}

Code search is a ubiquitous activity for developers, including data scientists. 
Sim \etal~\cite{sim2011how} comprehensively compared several code search tools including Koders\footnote{https://en.wikipedia.org/wiki/Koders}, Google, Krugle\footnote{https://krugle.com/}, and SourceForge\footnote{https://sourceforge.net/}, with different sizes of search target and search motivation. 
Common code search engines usually index code based on API keywords and method signatures.
Researchers have also utilized other information in code, such as structures, application descriptions, and data flows, to enhance the traditional keyword-based search; examples include Examplar \cite{mcmillan2012exemplar}, CodeGenie \cite{lemos2007codegenie}, and Sourcerer \cite{linstead2008sourcerer}.    
However, none of the above techniques are designed to tailor searching code or EDA processes in computational notebooks, which have different characteristics compared to traditional code modules. 
For example, in EDA notebooks, code cells are organized much freely and method signatures are difficult to extract. 
The analysis of structures and semantics relies on clean, well-documented, and linearly-organized code modules, which are often not available in computational notebooks. 

Visual methods have also been employed to understand the functional and structural components of code. 
Hoffswell \etal~\cite{hoffswell2018augmenting} proposed an in-situ visualization to summarize variable values and distributions.
Graph visualization is another popular way of presenting code, such as the dependency between variables and methods \cite{seider2016visualizing,balmas2004displaying}.
In the case of notebooks, Albireo \cite{wenskovitch2019albireo} uses a force-directed graph to display relationships among code cells and markdown cells.
TRACTUS \cite{subramanian2020tractus} employs a tree structure to reveal a hypothesis-driven analysis process. 
However, these systems focus on visualizing one single notebook, and do not provide the capabilities for searching or browsing a search result of multiple notebooks.
Lodestar \cite{raghunandan2021lodestar} uses a graph to model analysis steps in notebooks and provides recommendations for the next steps based on a semi-automatic analysis on a small corpus of about 6,000 notebooks. While this is similar to the API recommendation in our approach, it does not offer the in-situ search and exploration of code segments as we do, or thoroughly analyze high-level EDA patterns from a large enough corpus. 
Another similar work is NBSearch \cite{li2021nbsearch}, which supports semantic code search in a notebook collection and the exploration of resulting notebooks. But their search method is based on the code cell level, which is not effective in supporting EDA where higher-level blocks of code (containing multiple code cells) need to be retrieved and explored.  

There are also visualization techniques specially designed for presenting search results.
For example, Feng \etal's study \cite{feng2018effects} examined users' search behaviors in web visualization. 
Wilson \etal~\cite{wilson2010} advocated that web search interfaces should emphasize exploration. 
One way of presenting search results is based on a linear ranked list (\eg, Google). 
TileBars \cite{hearst1995tilebars} shows a colored bar next to each list item for the document length and term frequency. 
uRank \cite{sciascio2016rank} provides on-the-fly search results refinement and reorganization as a user's needs evolve.
Another way is to leverage a 2D space to present items with various layout methods. 
An energy model has been proposed to place text snippets of search results with minimal overlap \cite{gomeznieto2014similarity}. 
Space-filling techniques (\eg, treemaps) have also been used for browsing searching results~\cite{clarkson2009resultmaps,kleiman2015dynamicmaps}. 
Further, WebSearchViz \cite{nguyen2006novel}, Topic-Relevance Map \cite{peltonen2017topic}, and RankSpiral \cite{spoerri2004rankspiral} employ a circular (or spiral) layout, where the distance to the center represents a document's relevance, and the section of the circle denotes a specific topic or cluster. 
However, the above techniques focus on visualizing searched documents, rather than code or notebooks that have a different set of characteristics such as the presence of cells, variables, API calls, outputs, etc. 
The interactive visualization in \name{}, by contrast, displays EDA operations, code blocks, and the relationship between the searched code and other irrelevant code in notebooks.

\section{EDA in Computational Notebooks}

While EDA is not a new concept \cite{tukey1977exploratory}, topics on EDA with computational notebooks have recently gained much popularity in both industry and academia. 
In this section, to characterize EDA processes in notebooks, we first describe how we curated a corpus of EDA notebooks, which is the testbed of this work. We then report a formative study with professional data scientists as well as a quantitative analysis of the corpus regarding the EDA processes in notebooks.  

\subsection{Data Collection} \label{sec:dataset}

Rule \etal~\cite{rule2018exploration} provided a large collection of public Jupyter Notebooks scraped from GitHub. 
However, this corpus is noisy containing all kinds of notebooks with diverse goals, and it does not contain sufficient or clean metadata to determine whether a notebook is performing EDA or not.
Therefore, we curated a new corpus of notebooks by crawling high-quality submissions from Kaggle competitions. 
Each Kaggle competition features a data challenge, invites data scientists around the world to explore the data and build models to solve the problem, and evaluates their submissions with both automated testing and community feedback.
As the setup is more controlled, the notebooks tend to have a better quality as well as are well-formatted and well-documented. 
Also, as everyone works towards a single goal in a competition, it allows us to capture different approaches that data scientists used. 

We used the MetaKaggle\footnote{https://www.kaggle.com/kaggle/meta-kaggle} dataset as our entry point for getting access to Kaggle competitions. 
We selected competitions with the tags of ``featured,'' ``research,'' ``recruitment,'' and ``playground'' because the challenges are normally more open-ended, thus containing more EDA notebooks, and participated by expert data scientists, thus having higher-quality submissions. 
In total, we selected 281 competitions from Kaggle. 
For each selected competition, we filtered the notebooks by their ranks in accuracy on Kaggle (\ie, selecting the top 10\% of the total submissions). 
We also included notebooks that are never submitted but have a high number of upvotes and reviews, which are normally guide notebooks (\eg, written by the winners for explaining summarizing their methods aftermath).
In the end, we obtained a total of 38,581 notebooks, consisting of 856,941 code cells and 303,041 markdown cells. The median length of the notebooks is 22 code cells 75\% of all notebooks contain 39 cells at most. 

\subsection{Formative Study} \label{sec:formative-study}

While there exist many empirical and qualitative studies about data scientists' behaviors and workflows \cite{batch2018interactivea,kandel2012enterprise,subramanian2020tractus,li2021nbsearch,wang2019how}, many of them do not focus especially on EDA in computational notebooks. 
To better understand such EDA processes, we conducted semi-structured interviews with two professional data scientists (referred to as E1 and E2 below), recruited through the word of mouth.
Both experts have PhD degrees in computer science and have worked for three or more years as data scientists in large IT companies. Their job responsibilities include discovering business insights into customer data, deriving models and metrics for products, and creating visualization and reports, which consist of many EDA tasks with Jupyter Notebooks.   
The semi-structured interview included questions and discussion points on: how our interviewees perform EDA in their daily work, how they manage EDA with computational notebooks, what the key steps are in an EDA process, and what drives their decisions during EDA.
After, we used Otter.ai\footnote{https://otter.ai/} to transcribe the audio recordings of the interviews. Two authors independently coded the transcripts and then formed an affinity diagram together to discover themes and insights in the results. 
Our results are described as follows, which also confirm many observations from the literature.

\textbf{R1: Data scientists manage their EDA processes in semantic code blocks.}
As the nature of EDA is highly dynamic and uncertain, it is often challenging for data scientists to keep track of their analysis \cite{behrens1997principles,zhao2020chartseer}.
When using notebooks, they strive to maintain an analysis provenance, even for actions leading to dead ends \cite{rule2018exploration}. 
To mitigate the chance of getting lost and preserve the provenance, data scientists often organize their code into blocks and use these as checkpoints for navigation later, where each block represents one meaningful step in the EDA (\eg, loading data) \cite{subramanian2020tractus}.  
Our experts echoed the same behavior, where we refer to the code blocks as \textit{EDA blocks}. 
E2 said that a code block could contain one or multiple small code cells, but sometimes several semantic blocks are placed in \qt{a giant code cell.} 
E1 also mentioned \qt{I usually put all my preprocessing code and plotting code together, in one or several cells [...] I sometimes use PowerPoint slides to record the results of a section of code.}

\textbf{R2: A canonical EDA process generally contains a sequence of different semantic blocks.}
Several prior studies have investigated the general workflow of data science. 
Kandel \etal~\cite{kandel2012enterprise} identified five stages in enterprise data analysis and visualization, including discovery, wrangling, profiling, modeling, and reporting. 
Similarly, Batch \etal~\cite{batch2018interactivea} discovered four main elements in EDA based on context inquiries, including input (\eg, question), process (\eg, select, filter), environment (\eg, GUI, programming), and output (\eg, visualization).  
Subramanian \etal~\cite{subramanian2020tractus} also found data scientists' exploration involves standard routines such as finding base code, cloning, contextualizing, and evaluating the result.
Our experts described very similar steps in their EDA practices, where we tried to identify the basic units of EDA in terms of coding tasks. 
In the end, we concluded with four different types of EDA blocks, together forming a canonical \textit{EDA sequence}. They include:        
(1) configuration and data preparation, 
(2) model exploration and development, 
(3) hypothesis verification and evaluation, and
(4) output examination and visualization. 
The four stages were also confirmed by both experts. In addition, E2 mentioned: \qt{The benefit of Jupyter Notebook is the outputs (visual or textual) are persistent. I could complete a stage and move forward without rerunning the previous code again.}  

\textbf{R3: A real-world EDA process is normally iterative and non-linear, guided by data scientists' rationale.}
Our participants stated that in practice EDA is far more complex than we thought. Because EDA often explores \qt{the unknowns,} E1 made an analogy to the design thinking or user-centered design process \cite{brown2009change}. 
She said that \qt{an actual EDA process is iterative, going through the steps (basic EDA blocks) again and again, and also non-linear, jumping from one step (EDA block) to another,} not necessarily following the canonical order. 
Thus, sometimes an entire EDA process can be modeled as a tree structure or even graph \cite{subramanian2020tractus}, where each path could form an EDA sequence. 
Moreover, E2 indicated that such a real-world EDA process is guided by the rationale formed by a data scientist's prior experience, dynamic text and visualization outputs, and most importantly external knowledge (\eg, searched examples online). 
\qt{When I want to apply some non-trivial methodology, I frequently look for samples online. The API docs are not very helpful in this case,} said by E1.
This resonates Subramanian \etal's findings \cite{subramanian2020tractus} and the fact that code search is critical in exploratory programming with notebooks \cite{li2021nbsearch,brandt2009two,zhang2018are}.

\subsection{Analysis of Computational Notebooks} \label{sec:notebook-analysis}

To further characterize EDA processes, we performed quantitative analysis on our corpus of EDA notebooks, guided by the obtained results from the formative study. 
As real-world EDA is highly iterative and non-linear, a notebook may contain several interleaving canonical EDA processes (R3). 
First, we aimed to decompose a complicated notebook into multiple EDA sequences. 
To do so, for each cell that produces outputs (such as visualizations, printout tables, etc.) in a notebook, we checked the variable and function call dependency of that cell by using a similar program slicing technique in code gathering tools \cite{head2019managing}. 
The rationale of employing such a method is that we assumed a canonical EDA sequence ends by examining staged results with visual or textual outputs (R2).
Thus, each sliced code segment is an executable script from the original notebook, which mostly starts from the first few cells of the notebook that perform environment setup and data preparation, and ends at a cell that produces some intermediate (or final) outputs. 

We obtained 236,501 EDA sequences from our corpus.
Note that the script in each EDA sequence may be composed of code from more than one cell, and a cell may be split into different EDA sequences. 
While data scientists tend to organize their code in semantic blocks, sometimes different steps of EDA are written in one cell (R1).
The above analysis breaks the original cell boundaries and builds based on code dependency, which allows for better capturing the semantics in code. 
However, we observed that in most of the cases, cells were ``preserved'' in our sliced programs, because many cells just contain small and atomic chunks of code or a single function definition.      

Moreover, we aimed to analyze the steps in our sliced EDA sequences. 
Inspired by the literature of text analysis, we employed topic modeling \cite{blei2003latent} to discover themes in the code. 
Intuitively, we treated each EDA sequence as a ``document'' and tried to identify its ``topic'' composition, where each ``topic'' governs code that is semantically coherent.  
However, different from natural language documents mostly containing English words, code may exhibit a larger vocabulary because variables can be freely named.
Thus, we only extracted the APIs from common data science toolkits (\eg, pandas, numpy, scipy, scikit-learn, and some Python built-in functions) as the ``words'' in ``documents.'' 
By analyzing the code dependency and structure, we extracted full API calls as our tokens; for example, we expanded the python builtin API \texttt{len} to \texttt{\_\_builtins\_\_.len}. This normalized different forms of calling an API in different notebooks, reducing non-necessary duplications. 
In total, we identified 19,453 unique API keywords, denoting the vocabulary of notebooks' code.

\begin{figure*}[tb]
    \centering
    \includegraphics[width=\linewidth]{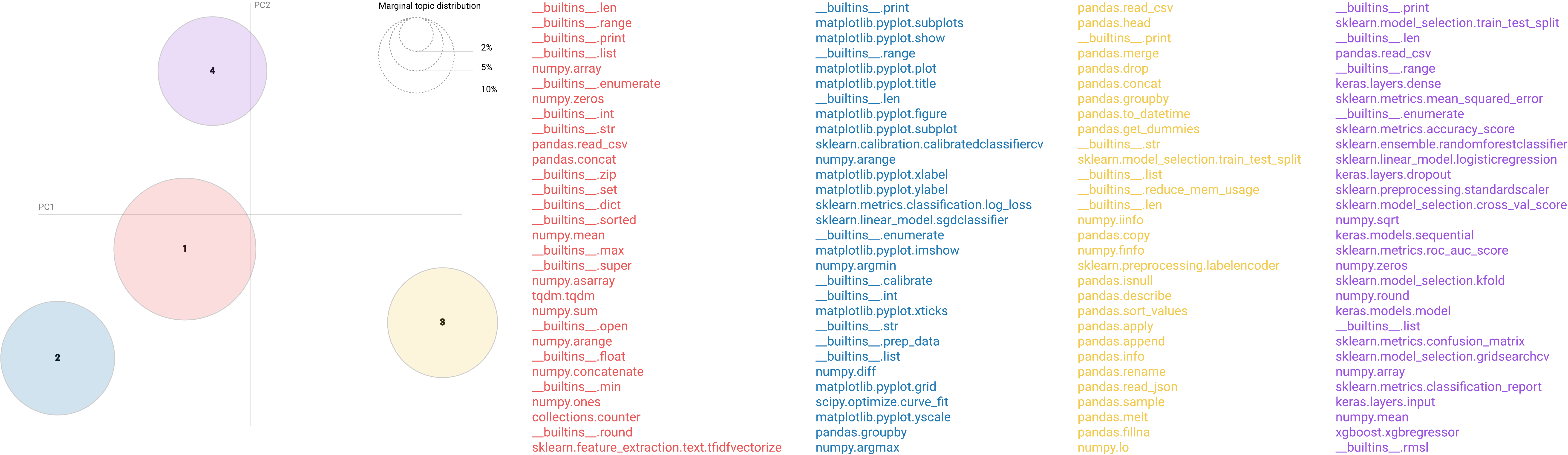}
    \vspace{-7mm}
    \caption{Topic analysis of the EDA sequences in our notebook corpus with GuidedLDA. The top 30 frequent keywords of each topic are shown in the corresponding color. The figure is generated using pyLDAvis and is in high resolution.}
    \label{fig:topics}
\end{figure*}

Next, we performed LDA \cite{blei2003latent}, a widely-used method in topic modeling, on EDA sequences represented by these keywords. 
To thoroughly explore the data, we varied the number of topics in the LDA input. 
When there are four topics, the code scripts are relatively separated; when there are five topics, two of them overlap a lot, which could be combined into one topic with many python built-in and common data science APIs. These observations were based on our exploration of the results with the pyLDAvis\footnote{https://github.com/bmabey/pyLDAvis} visualization. 
This confirms our formative study results that there exist four different EDA operation types (R2).
We then examined the keywords, which matched our expectations for the four types. 
To obtain more precise and descriptive topics, we further conducted GuidedLDA \cite{jagarlamudi2012incorporating} using some most salient keywords selected from our initial LDA results. 
As shown in \autoref{fig:topics}, the resulting four topics are relatively separated based on PCA (principal component analysis). 
From the top keywords shown on the side, we can roughly discover: Topic 1 (red) is about hypothesis verification and evaluation (mostly builtin and numpy APIs); Topic 2 (blue) is about output examination and visualization (mostly matplotlib APIs and \texttt{print}); Topic 3 (yellow) is about configuration and data preparation (mostly pandas APIs); and Topic 4 (purple) is about model exploration and development (mostly sklearn and keras APIs).

\section{\name{} Overview}
The curated corpus, formative study, and quantitative analysis of the notebooks have set the basis of the design of \name{}, which aims to facilitate open-ended EDA in computational notebook environments.
In this section, we first introduce the design goals of \name{}, and then an overview of the \name{}'s architecture, followed by a scenario of using the tool. 

\subsection{Design Goals} \label{sec:design-goals}
Based on our understanding of the challenges of EDA from the formative study and the literature, we distilled the following design goals to guide the development of \name.

\textbf{G1: Provide suitable EDA examples in context.}
Because real-world data problems are often vague and ill-defined, data scientists usually search for existing EDA notebooks online to learn how others address similar issues \cite{head2020composing,subramanian2020tractus,li2021nbsearch}. Our E1 and E2 confirmed this as well (\textbf{R3}).
Code search is an essential activity for almost all developers beyond just data scientists in conducting EDA \cite{brandt2009two,robillard2010field,sim2011how,zhang2018are}. 
Therefore, the system should be able to retrieve appropriate EDA examples from the widely available public EDA notebooks online, which is necessary for data scientists, especially novices, to get up speed with their EDA.
However, currently data scientists still use external tools such as Google Search to achieve this task. 
Thus, the retrieval support needs to be in place, within their programming environment (\eg, Jupyter), and in-situ, closely associated with the code scripts they are working on.    

\textbf{G2: Facilitate exploration of example EDA processes.} 
As mentioned earlier, a real-world EDA process exhibited in a notebook is often interleaving and non-linear (\textbf{R3}), although data scientists organize their code in semantic blocks (\textbf{R1}).
Even a set of suitable notebooks are retrieved for the context, it is challenging for a data scientist to quickly comprehend others' EDA processes and make use of the search results \cite{li2021nbsearch,head2019managing}. 
Currently, data scientists can only rely on an embedded renderer (\eg, on GitHub or Kaggle) to view notebooks and try to decipher others' code which may also contain a lot of irrelevant information.  
Thus, the system should extract the relevant parts from an entire notebook and present those parts first in a complete EDA sequence (\textbf{R2}). Further, the system should support interactive exploration of the complicated EDA processes reflected in the retrieved notebooks.

\textbf{G3: Offer suggestion for subsequent EDA actions.}
While with examples notebooks data scientists can obtain knowledge from others' code such as the APIs used, the number of examples they can view during their EDA is limited. 
Being able to provide some suggestions about subsequent exploration steps can be significantly helpful, not only in data manipulation \cite{barel2020automatically,yan2020autosuggest} but also in visualization generation \cite{wongsuphasawat2016voyager,wongsuphasawat2017voyager,mackinlay2007show}.
This is essential for novices to learn and get familiar with a large number of methods/APIs available in data science toolkits (\eg, pandas) \cite{guo2012burrito,head2020composing}, thus helping them form rationale for the next steps (\textbf{R3}).
While completely automating the whole EDA process without any limitation is not possible, the system should offer some level of suggestion, also in-situ, such as the operations that are likely to use next based on the current code sequence (\textbf{R2}). 
This allows data scientists to effectively make decisions for carrying out their EDA.     

The existing practice for data scientists to search for examples during EDA is based on frequently switching between the Jupyter interface and external search engines such as Google and StackOverflow. 
Compared to this, the above design goals have outlined the demand for more context-based code search, recommendation, and exploration and more integrated data science workflow with computational notebooks. 
However, our goal here is not to replace the existing practice, but to investigate effective means of facilitating EDA with in-situ support during the EDA process.

\subsection{System Architecture}

\begin{figure}[tp]
    \centering
    \includegraphics[width=\textwidth]{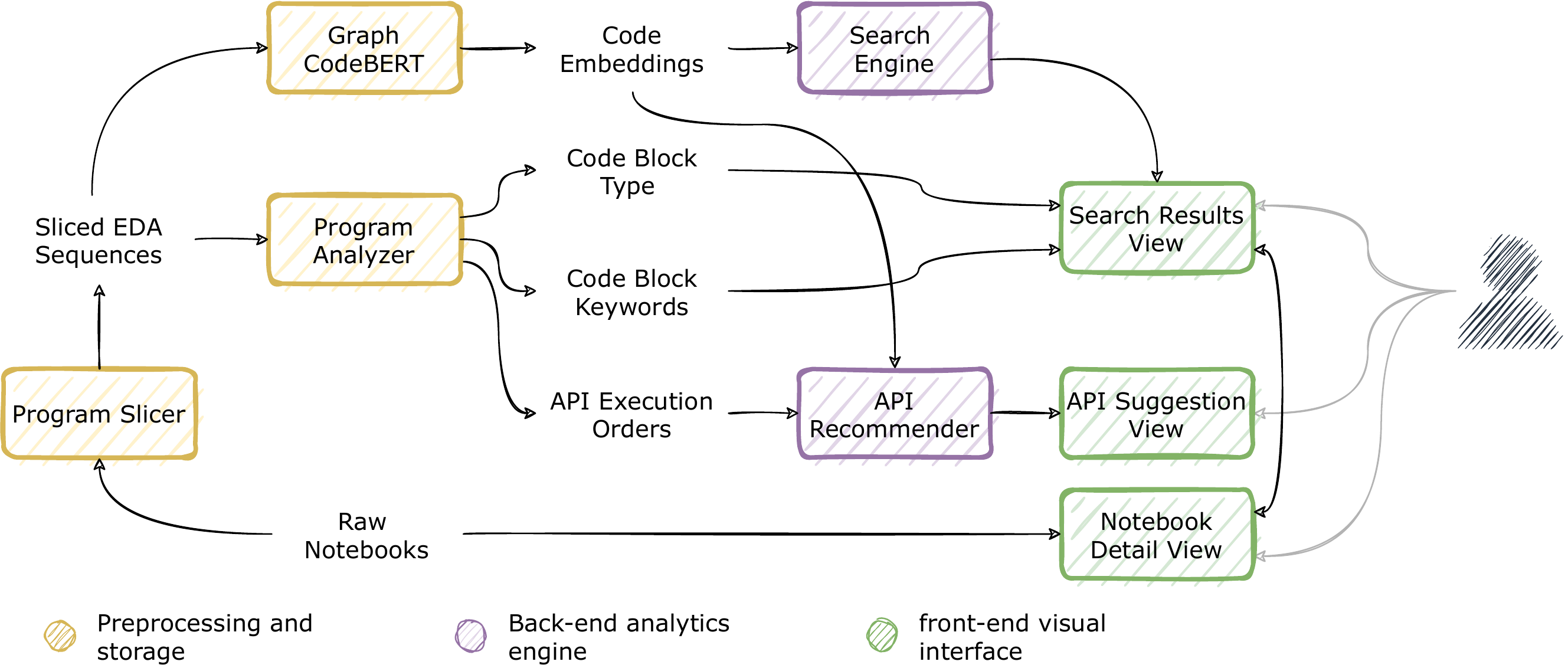}
    \vspace{-4mm}
    \caption{\name{} system architecture, which consists of three components: (1) a Data Preprocessing and Storage module, (2) a back-end Analytics Engine, and (3) a front-end Visual Interface.}
    \label{fig:system-overview}
\end{figure}

Following the above design goals, we developed \name, an interactive and visual tool that facilitates EDA with code search, exploration, and recommendation based on existing notebook repositories. 
\name{} is embedded within the Jupyter Notebook environment as an extension for seamless access to these functionalities.
As shown in \autoref{fig:system-overview}, the \name{} system consists of three components: (1) a \textit{Data Preprocessing and Storage} module, (2) a back-end \textit{Analytics Engine}, and (3) a front-end \textit{Visual Interface}. 
Details about these components will be introduced later.   

In the Data Processing and Storage module, from the raw corpus of EDA notebooks (see \autoref{sec:dataset}), a \textit{Program Slicer} disentangles EDA sequences from the original notebooks, as described in \autoref{sec:notebook-analysis}. 
Each sliced EDA sequence, which is an executable script, is fed to a \textit{Program Analyzer} (see \autoref{sec:program-analyzer}) that extracts three types of information: (1) used APIs or methods in the order of execution, (2) descriptive keywords for code blocks, and (3) EDA types for code blocks, based on the LDA topic analysis in \autoref{sec:notebook-analysis}.    
The sliced EDA sequences are also used for fine-tuning the GraphCodeBERT \cite{guo2020graphcodebert}, a pre-trained code representation deep learning model, for our two downstream tasks: EDA sequence search and next API prediction. GraphCodeBERT generates a set of code embeddings, which is stored in a database, along with other computed information above, for later use in other components of the system.

The back-end Analytics Engine includes two key components, which work interactively with the front-end Visual Interface that contains three main views. 
First, a \textit{Search Engine} (see \autoref{sec:search-engine}) leverages the code embeddings to retrieve potentially useful EDA examples based on a code query from the front-end (\ie, relevant code in the notebook based on the current working cell), and all the examples are then summarized in a \textit{Search Results View} (\autoref{fig:system-interface}-b; see \autoref{sec:search-view}) with a novel visualization (G1, G2). 
Second, an \textit{API Recommender} (see \autoref{sec:api-recommender}), also built upon the GraphCodeBERT model, utilizes the code embeddings and extracted API execution orders to predict APIs that are most likely to use next.
The recommended results, which are obtained from a code query constructed in a similar way as above, are displayed in the \textit{API Suggestion View} (\autoref{fig:system-interface}-d; see \autoref{sec:suggestion-view}) on the front-end (G3).   
Finally, the front-end \textit{Notebook Detail View} (\autoref{fig:system-interface}-c; see \autoref{sec:detail-view}) allows for viewing the detailed code of each searched EDA example within the context of the original notebook that it was extracted from (G2).

\subsection{Usage Scenario}

\begin{figure*}[tb]
    \centering
    \includegraphics[width=\linewidth]{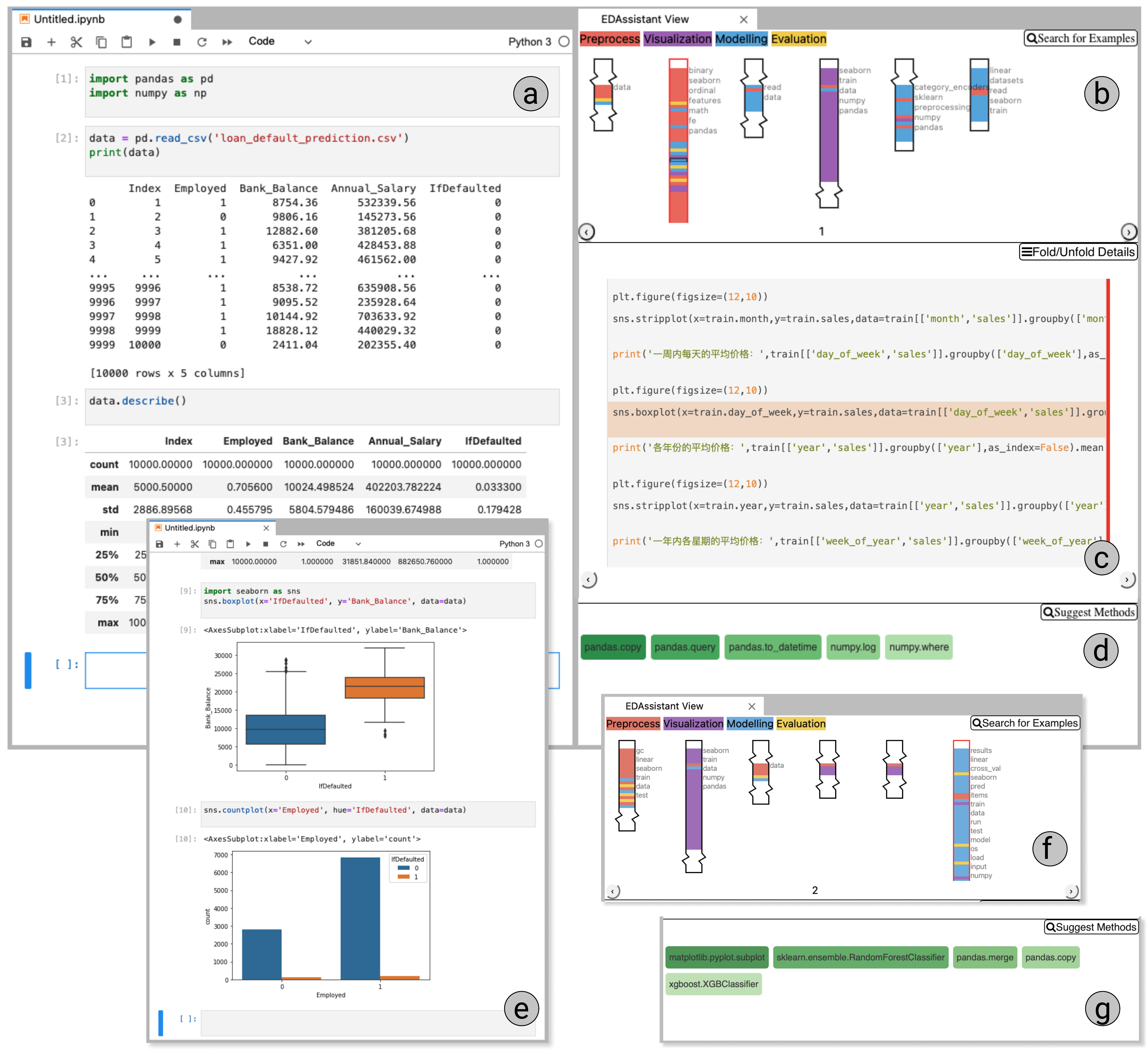}
    \vspace{-7mm}
    \caption{A data scientist is conducting EDA on a bank loan default dataset with \name{}, which is a JupyterLab extension to offer situated EDA support with three interactively coordinated views for Search Results (b), Notebook Detail (c), and API Suggestion (d).}
    \label{fig:system-interface}
\end{figure*}

In this section, we demonstrate the basic usage of \name{} with a simple scenario.
Suppose that Alex is a junior data scientist who just starts his job at a bank, and he is asked to find some insights from some customer loan default records. 
After loading the dataset and printing out some portions of the data table in JupyterLab, Alex gets stuck on how to proceed with his EDA. Thus, he launches \name{} and places the panel beside his notebook (\autoref{fig:system-interface}-a). 

He clicks ``Search for Examples'' and within seconds, \name{} returns a rank list of EDA sequences based on his current notebook and working code cell, organized horizontally in the Search Results View  (\autoref{fig:system-interface}-b). 
Each example EDA sequence is visualized as colored strips stacked together, where each strip represents a continuous code block in the searched notebook and the color indicates the EDA operation type (\ie, preprocessing, modeling, visualization, and evaluation, as shown in the legend). The white strips (spaces) indicate other less important code in the notebooks.
He has an intuition that some visualization of the data is needed before building any models, so he clicks on the purple parts of the searched EDA sequences. This displays the corresponding code in the Notebook Detail View and jumps to the selected line of code (\autoref{fig:system-interface}-c).

Similarly, he browses a few top-ranked examples; and he finds out that the methods \texttt{countplot} and \texttt{boxplot} from the seaborn library are used often for initial data visualization. 
By toggling the ``Fold/Unfold Details'' button, he is able to switch his focus between the relevant code scripts and the whole example notebook. He can also browse the markdown cells and in-line comments available in the examples to get some context about how the methods are used and why.  
Some basic data manipulation APIs are recommended as well (with importance encoded by color transparency), when Alex clicks ``Suggest Methods'' (\autoref{fig:system-interface}-d). 
Following the API usages in these examples, Alex easily swaps some parameters and creates a few charts to view several basic characteristics of his data, such as the distributions of \texttt{Bank\_Balance} over \texttt{IfDefaulted} (\autoref{fig:system-interface}-e). 

Now Alex identifies some relationships between \texttt{IfDefaulted} and other data variables. So he wants to build some machine learning models to predict the loan default state.  
Again, he has no idea what technique to use. Thus, he clicks ``Suggest Methods'' which triggers the API recommendation and returns a list of potentially useful methods next, based on his current code. 
Alex finds that the second highest ranked method \texttt{RandomForestClassifier} from scikit-learn could be an interesting technique to try (\autoref{fig:system-interface}-g). 
Hovering over the API provides a brief method summary from its official documentation, and clicking it directs Alex to the documentation page online.  
With this in mind, Alex copies and pastes the API call signature to his notebooks, and then clicks ``Search for Examples'' again. A different set of examples is retrieved based on his updated code (\autoref{fig:system-interface}-f), which indeed contains some sample usage of \texttt{RandomForestClassifier}. 
Thus, Alex decides to follow the examples to apply this classifier to his prediction problem.

\section{Notebook Analytics in \name{}}

In this section, we describe the data processing and analysis in \name{}, including how we prepared the data, built the search function, and constructed the API recommender.

\subsection{Analyzing Sliced EDA Sequences} \label{sec:program-analyzer}

The preprocessing and storage module of \name{} is based on the previous explorative analysis described in \autoref{sec:notebook-analysis}. 
As shown in \autoref{fig:system-overview}, the raw notebooks are sliced into EDA sequences by the Program Slicer based on the approach in code gathering tools \cite{head2019managing}. 
The sliced scripts are then processed by the Program Analyzer that produces the following three kinds of outputs for future use in \name.

\begin{itemize}    
    \item \textbf{Code EDA operations.} 
    \autoref{sec:notebook-analysis} describes our exploration of the EDA operation types based on topic modeling, which detects four ``topics'' in code: (1) configuration and data preparation, (2) model exploration and development, (3) hypothesis verification and evaluation, and (4) output examination and visualization.   
    Leveraging the learned topical keywords and their probability distributions within each topic, the main EDA operation type can be identified for any given code block.
    However, it is challenging to define the boundaries of semantic code blocks, as data scientists sometimes split them into different code cells or just write a giant code cell for several goals (see \autoref{sec:formative-study}). As an initial step, we relied on the code cells from the original notebooks to determine code blocks. That is, if the lines of code are from the same code cell originally, they stay in the same code block in the sliced script. 
    Thus, the Program Analyzer generates a sequence of high-level EDA operations based on the order of the code blocks and their contents.

    \item \textbf{API execution orders.} 
    \autoref{sec:notebook-analysis} also describes how we extracted the keywords for the topic modeling, which include the APIs from common data science toolkits such as pandas.  
    The Program Analyzer keeps this information and outputs the API or method execution orders of each sliced script based on its parsed abstract syntax tree. 

    \item \textbf{Code keywords.} 
    Further, the Program Analyzer outputs a set of descriptive keywords for each of the code blocks in the sliced EDA sequences, which helps summarize the gist of the code. 
    We utilized a simple TF-IDF approach \cite{ramos2003using} in document retrieval, by viewing each EDA sequence as a ``document'' and each extracted APIs or methods in code as ``words.'' 
    The keywords with top TF-IDF scores are treated more informative for describing the content of the code.  
\end{itemize}


\subsection{Retrieving Example EDA Sequences} \label{sec:search-engine}
As shown in \autoref{fig:system-overview}, we fine-tuned GraphCodeBERT \cite{guo2020graphcodebert} to build the Search Engine in \name{}, which can retrieve suitable EDA examples from our notebook corpus, dynamically based on the code that a data scientist currently works on (\textbf{G1}).
Pre-trained models such as BERT \cite{devlin2018bert} have shown significant advantages in natural language processing (NLP) tasks. The pre-trained models are first trained on a large unsupervised corpus to generate latent representations of the text (\ie, embeddings), and then fine-tuned on downstream tasks.
GraphCodeBERT is the state-of-the-art model for programming languages based on similar structures of BERT and CodeBERT \cite{feng2020codebert}, which is comprised of an encoder (that converts code into embeddings) and a decoder (that converts embeddings into code). 
We chose GraphCodeBERT because it also leverages the data flow graph in programs for training and achieves the best performance for many downstream tasks such as code clone detection and natural language code search.

In our development, we followed the same approach in their code search downstream task to train GraphCodeBERT with massive EDA sequences, each being a series of code blocks as described in \autoref{sec:program-analyzer}.
Intuitively, each code block is like a ``word'' in a ``sentence'' that is the whole EDA sequence, and our goal is to obtain the embeddings of the EDA sequences for our Search Engine. 
Thus, guided by SentenceBERT \cite{reimers2019sentence}, we added a mean pooling layer to obtain the ``sentence'' embeddings, which has shown effectiveness in their experiments.  
We used 60\% of the data for training, 20\% for validation, and 20\% for testing.   
After the training process, each EDA sequence can be represented by a 768-dimensional vector, and an encoder that can transform any EDA sequences into embeddings is obtained.
Therefore, during the inference time, the Search Engine takes an input EDA sequence, uses the encoder to get its embedding, and leverages the embedding to find relevant EDA sequences in our corpus based on cosine similarity. 
These related examples could help data scientists get inspiration on what to do next in their own EDA.   

\begin{figure}[tb]
    \centering
    \includegraphics[width=0.6\textwidth]{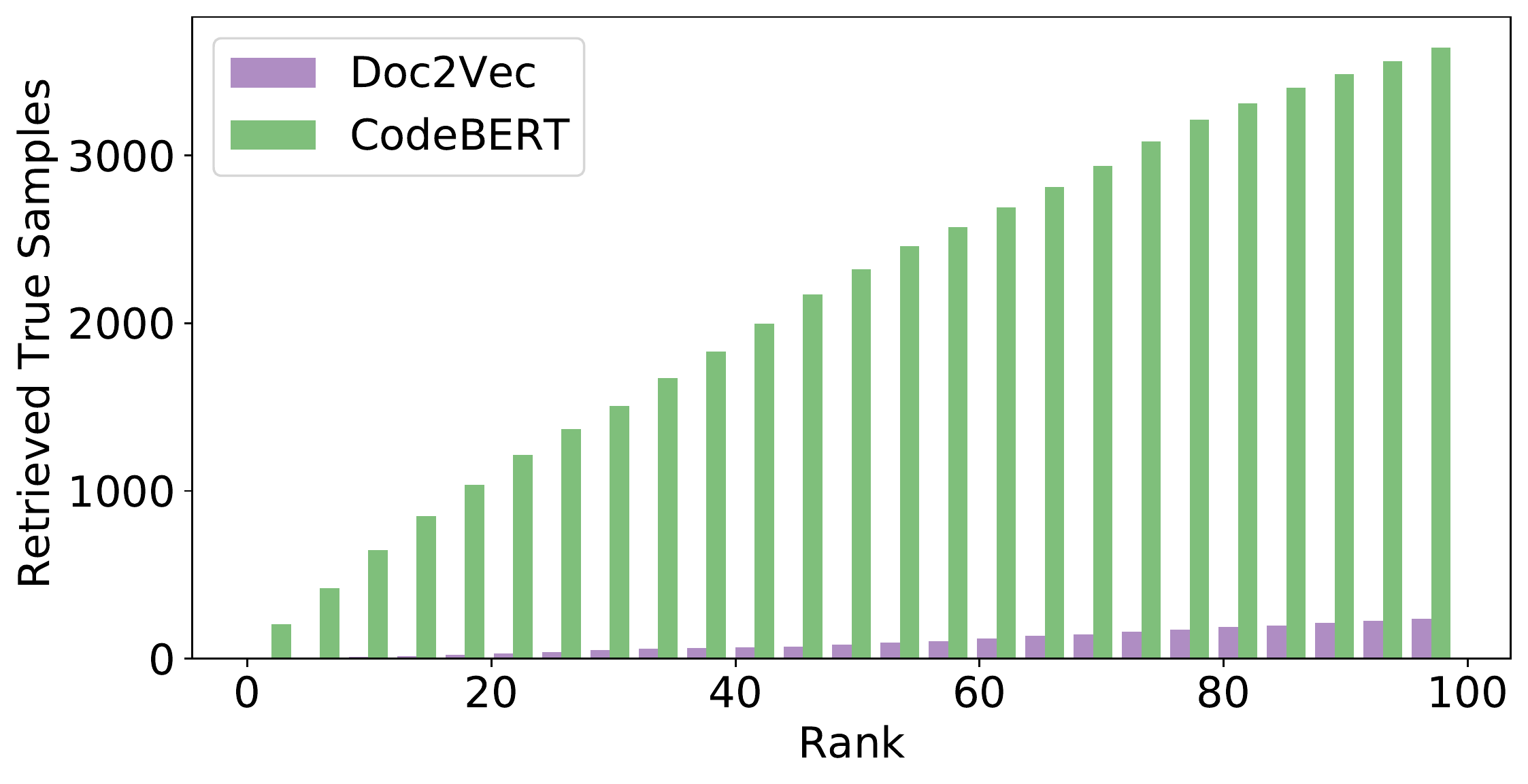}
    \vspace{-5mm}
    \caption{Comparison of GraphCodeBERT and Doc2Vec on the numbers of true samples retrieved in the top-$k$ search results.}
    \label{fig:search-eval}
\end{figure}

To investigate how well GraphCodeBERT works in our situation, we compared it with a baseline, Doc2Vec \cite{le2014distributed}, which was trained to generate embeddings with the same size. Doc2Vec is a classic NLP method for unsupervised document representation learning, not based on neural networks. 
We chose this method because the experiments in GraphCodeBERT already compared the model with a couple of state-of-the-art neural network based models.
The two models were compared using simulated code searching tasks. 
For each EDA sequence in our test dataset, we used its first $N$ code blocks as the query and checked the rank of the original full EDA sequence (\ie, ground truth) in the retrieved results. 
Thus, we generated $N-1$ queries for each EDA sequence to perform the experiments, where $N$ is the length of the sequence.
\autoref{fig:search-eval} shows the counts when the ground truth falls within the top-$k$ items of the search results (until top 100).
We can see that GraphCodeBERT returns significantly more correct EDA sequences than those by Doc2Vec for any given rank; Doc2Vec barely retrieves any true samples within top-$20$. 


\subsection{Recommending Potentially Useful APIs} \label{sec:api-recommender}
Besides providing suitable examples to inspire data scientists on how to perform their current EDA, \name{} recommends the potential APIs to use next via the API Recommender (\textbf{G3}). 
It is built upon the same architecture as the Search Engine model using GraphCodeBERT. Specifically, a dense layer is added after the last EDA sequence embedding layer to predict a one-hot vector (in 19,453-dimension) representing the next API to use over all the possible APIs (\ie, our vocabulary).
During training, in addition to the EDA sequences as the input, the extracted API execution orders (see \autoref{sec:program-analyzer}) were used as ground truth for the target prediction.    
This design allows the Search Engine and API Recommender models to share the same architecture and achieve two tasks simultaneously with one training process, significantly increasing the efficiency.

Similarly, we evaluated the API Recommender by comparing it with two different baselines. 
The first is the Search Engine itself, where the APIs of the top-ranked EDA sequence in search results were treated as recommendations. This baseline helps us investigate whether the added dense layer is necessary. 
The second is based on the Doc2Vec model, where a similar approach was used to get recommended APIs (\ie, extracted from the search results).
However, the GraphCodeBERT with a dense layer predicts a probability distribution over all possible APIs, rather than a set of APIs like the two baselines. We thus used $\log(p_i)>0.5$ as a threshold to select the APIs where $p_i$ is the probability of API$_i$.
We used a similar experimental setup to \ref{sec:search-engine} to simulate the processes of getting API recommendations.
For each EDA sequence in the corpus, we used its first $1$ to first $N-1$ code blocks to create $N-1$ queries and compared the predicted APIs with ground truth APIs.

\begin{figure}[tb]
    \centering
    \includegraphics[width=0.6\textwidth]{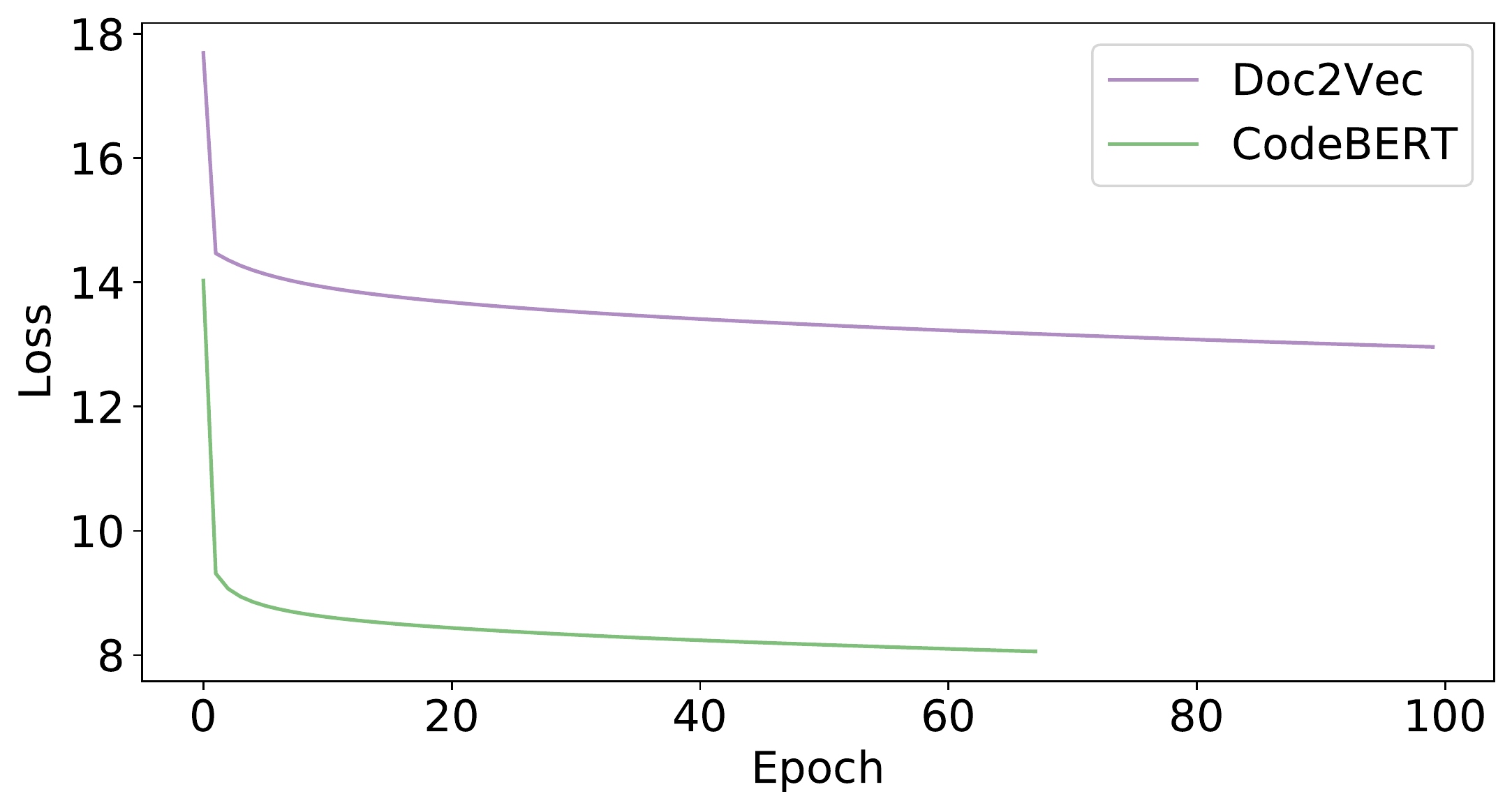}
    \vspace{-5mm}
    \caption{Comparison of GraphCodeBERT and Doc2Vec on training losses for API recommendation. }
    \label{fig:recommender-loss}
\end{figure}

Given these predicted sets of next APIs, we computed the accuracy of the models by averaging the number of correctly predicted APIs divided by the actual number of ground truth APIs in each query.
However, the accuracy measure does not consider the size of predicted sets, because when the size of a predicted set is larger, more ground truth APIs are likely in it. 
We thus computed the IOU (intersection over union) between the predicted sets and the ground truth sets, so larger predicted sets get some penalty. 
As shown in \autoref{tab:api-recommendation}, the results indicate that the API Recommender model is comparable with Doc2Vec, but significantly outperforms the bare Search Engine model without the dense layer.     
Given that GraphCodeBERT significantly outperforms Doc2Vec for the code searching tasks (see \autoref{sec:search-engine}), it is thus still valuable to use the GraphCodeBERT-based models in our cases due to the efficiency in sharing the code embeddings. 
Further, as shown in \autoref{fig:recommender-loss}, GraphCodeBERT converged much faster during training than Doc2Vec, which could be more applicable in practice for larger training datasets.

\begin{table}[tb]
    \centering
    \small
    \caption{Comparison of different API recommendation models with accuracy and IOU.}
    \label{tab:api-recommendation}
    \vspace{-3mm}
    \begin{tabular}{lcc}
        \toprule
        &  Accuracy & IOU   \\
        \midrule
        GraphCodeBERT with a dense layer (API Recommender) & 0.507 & 0.399  \\
        GraphCodeBERT only (Search Engine) & 0.146 & 0.079  \\
        Doc2Vec & 0.491 & 0.389 \\
        \bottomrule
    \end{tabular}
\end{table}

\section{User Interface of \name{}}

In this section, we introduce the front-end visual interface of \name{} (\autoref{fig:system-overview}) which is developed as an extension of JupyterLab. It can be accessed on a panel next to the main notebook panel and contains three interactively coordinated views (\autoref{fig:system-interface}): the Search Results View, Notebook Detail View, and API Suggestion View.  

\subsection{Visualizing Searched EDA Sequences} \label{sec:search-view}

At any time during the EDA, a data scientist can initiate a search for useful EDA sequences to guide their analysis, by clicking the ``Search for Examples'' button on the top panel (\textbf{G1}). \name{} analyzes the currently active code cell in the notebook, extracts dependencies from other cells, and constructs a sequence of code blocks as the query for the Search Engine. 
Then, \name{} displays the retrieved EDA sequences in the Search Results View (\autoref{fig:system-interface}-b), with a novel visualization, called \textit{DNA plot}. 
The visual design resembles the chromosomes on DNAs, and the retrieved EDA sequences are displayed from left to right based on their search ranks.
Prior work indicates that visualization of search results and code, rather than textual ranked lists, is necessary for users to better understand and explore the returned items (\eg, \cite{li2021nbsearch,nguyen2006novel,peltonen2017topic,hearst1995tilebars}).
Our visualization here can help data scientists get an overview of the search results and thus make decisions to explore specific EDA sequences (\textbf{G2}). 

Fundamentally, each EDA sequence is extracted from a real-world notebook, and the notebook is also important to understand the EDA sequence. However, exposing too much unnecessary information could overwhelm users.    
The DNA plot encodes an example EDA sequence in the context of its original notebook, with horizontal colored strips indicating each code block inside the sequence and white space indicating those not belonging to the sequence (\autoref{fig:system-interface}-b). 
The color of the strips represents the EDA operation type of the code block.
As the original notebook might be lengthy, a ``folded'' visual metaphor is used to indicate many consecutive code blocks that are not in the retrieved EDA sequence. 
Hovering over the strip or the white space initiates a tooltip of the actual code it represents.
Further, a set of code keywords (one of the outputs of the Program Analyzer; see \autoref{fig:system-overview}) are shown beside the DNA plot to provide more contextual information (\autoref{fig:system-interface}-b).
The DNA plot design facilitates data scientists with showing approximately where the EDA sequence is located in the original notebook and what operations the EDA does in general.

\subsection{Exploring EDA Sequences in Context} \label{sec:detail-view}

Based on the search results, the Notebook Detail View of \name{} allows a data scientist to browse the actual code in an EDA sequence in detail by clicking the corresponding DNA plot (\textbf{G2}). Moreover, clicking a color strip navigates the view to the specific lines of code it represents (\autoref{fig:system-interface}-c). 
The code is presented in a similar visual fashion to the JupyterLab user interface for ease of learning. 
Initially, only the code belonging to the selected EDA sequence is shown in the Notebook Detail View.
As each EDA sequence is extracted based on variable dependencies, it is a complete and executable script that is normally sufficient for the data scientist to understand the searched examples.  
However, clicking the ``Fold/Unfold Details'' button on the middle panel toggles other cells in the original notebook, which provides more context on demand. 
To distinguish the cells within the EDA sequence from other cells, a small red vertical bar is shown on the right for each cell in the sequence and the current select line of code is also highlighted. 
This design avoids presenting too much information at once, increasing the efficiency of comprehending the code, leveraging the useful knowledge, and browsing the search results. 

\subsection{Discovering Subsequent APIs to Use} \label{sec:suggestion-view}

In addition to the searched EDA sequences that allow a data scientist to learn from the examples and carry out their own EDA, \name{} recommends common APIs that can be potentially used next, also based on the current coding context. 
The data scientist can initiate the recommendation by clicking the ``Suggest Methods'' button in the API Suggestion View (\autoref{fig:system-interface}-d).
This triggers the API Recommender of \name{} and the returned APIs are displayed as tags with the color transparency indicating the probability. The darker the color is, the more likely the system thinks that API is useful.
Further, clicking the tag links to the online documentation of the corresponding API, which helps the data scientist learn about its usage. 
While only a set of common data science toolkits are considered now in \name, it is easy to integrate more APIs in the future. 
\section{User Study}

To assess the effectiveness and usefulness of \name{} in supporting EDA, we conducted a user study by investigating our tool and a baseline approach that resembles the setup of data scientists' current practice (\ie, using a separated search tool, Google Search, while conducting EDA in computational notebooks). 
However, we note that Google Search can access a much larger corpus of notebooks and knowledge base than ours. We do not want to compare \name{} with a customized search engine with only our dataset, because that might constrain the user experience and our investigation.
Our goal here is not to replace the existing search engines, but to understand the strengths and weaknesses of both approaches and seek opportunities for \name{} to be used in data scientists' workflows. 
The general purpose of the study is to explore users' experience of finding useful examples during EDA with their current practice and \name.     

\subsection{Participants}
We recruited 14 participants, 12 males and 2 females, aged 22--41 (\mean{27.8}, \sd{5.8}), through mailing lists at a local university and social media.
Their technical backgrounds included computer science and engineering. Of all the participants, two were with PhD degrees, seven with Master's degrees, and five with Bachelor's degrees. Further, four of them were professional data scientists and the rest were students.
We did a pre-screening for participants' experience in using Python, computational notebooks, and relevant data science toolkits.
We selected the participants who met the minimum technical standard for a novice or entry-level data scientist.
On a 1--7 Likert scale (from ``no experience'' to ``expert''), participants' self-reported technical knowledge was as the follows: Jupyter, \md{6}, \iqr{0.75}; Python, \md{6}, \iqr{0.75}; chart plotting libraries such as MatplotLib, \md{6}, \iqr{1}; machine learning libraries such as scikit-learn, \md{5}, \iqr{0}; data science libraries such as Pandas and numpy, \md{6}, \iqr{0.75}; as well as familiarity with visualization, \md{5}, \iqr{0}.

\subsection{Tasks and Design}
Two datasets were selected from the test set of our notebook corpus, which have similar sizes (\ie, the numbers of columns and rows) and complexity. Both were tabular datasets including categorical and numerical attributes.
One was about students' exam scores with their demographic information (\eg, gender, catered lunch or not), and the other was about customers' loan defaults with their personal data (\eg, income, employed or not).

We adopted a within-subjects design for our study. 
The task resembled a realistic open-ended EDA with a controlled structure, which contained two parts.
The first part was more constrained, in which participants were asked to plot charts on certain attributes. The purpose of this was to help participants get familiar with the dataset and warm up.
The second part was more open, in which participants were asked to investigate the data patterns in more depth by building models to predict certain values, cluster the data points, or classify the records. Participants were encouraged to plot charts during this exploration process.
Participants were instructed to search for examples freely only using the provided tool (\ie, Google Search or \name).
To make the study more trackable, they were only allowed to use the loaded libraries in a starter Jupyter notebook, which included pandas, numpy, scipy, matplotlib, seaborn, and scikit-learn. These libraries were common data science toolkits and were sufficient for the study tasks.
The order of presenting the study tools and the datasets was counter-balanced with a Latin square design across participants.

\subsection{Procedure}
The study was conducted using remote conferencing software, where participants used their own computers to access \name{} hosted on a server. 
At the beginning, they were introduced to the general background and procedure of the study. Then, each participant was asked to completed two EDA tasks as described above, one with each tool (\ie, Google Search or \name).
For \name, a brief video tutorial was provided right before the task and participants could ask any questions about the tool's functionalities.
As EDA is usually open-ended and there is no clear indication of completeness, we set a 20-minute limit for each task, where participants were encouraged to explore the data as much as possible.
There was no hard time limit for each part of the task but they got a reminder around 9 minutes. 
After each task, participants filled in a short questionnaire regarding their experience of using the tool.
In the end of the study, a semi-structured interview was conducted to collect participants' qualitative feedback about the two tools. 
The whole study lasted about one hour for each participant, and they received \$10 for remuneration.   

\subsection{Results and Analysis}
In this section, we report the results of our user study, which includes participants' task performance, subjective ratings on their experiences, and qualitative feedback to \name. Participants are referred as P\# in the following text.

\subsubsection{Quantitative Results}

\begin{table}[tb]
    \centering
    \small
    \caption{Comparison of participants' task performance with different indicators by \textsc{M} and \textsc{SD} (in parentheses). }
    \label{tab:task-performance}
    \vspace{-3mm}
    \begin{tabular}{lcc}
        \toprule
         &  Google Search & EDAssistant \\
        \midrule
        Completion Time (minutes) & 9.8 (2.5) & 12.9 (3.7)\\
        Number of Lines Written & 16.1 (9.2) & 17 (15.5) \\ 
        Number of Cells Created & 7.1 (1.8) & 6.4 (1.3) \\
        Number of Charts Generated & 1.7 (0.8) & 1.8 (0.4) \\
        Number of Searches Performed & 4.2 (1.8) & 3.6 (1.9)\\
        Number of Users Built Models (out of 14) & 6 & 10 \\ 
        \bottomrule
    \end{tabular}
\end{table}

\autoref{tab:task-performance} shows different task performance metrics of Google Search and \name{}.
There was no significant difference in time between the tools for participants to complete the two study tasks and reach the results that they were satisfied with. 
While participants with \name{} took longer time on average, as EDA often has vague goals, task completion time is less of an indicator. But this is still encouraging because the new visualization and interface of \name{} that require further familiarity did not significantly slow the participants down. 

The ending states of participants' notebooks at the end of the study were similar between the two conditions, in terms of the number lines, code cells, and charts created. The reason that \name{} has a high variance for the number of lines is because P9 wrote 68 lines in their exploration.
These results demonstrate that both the tools could lead participants to reasonable EDA for our tasks, while with Google Search, participants have the advantage of getting access to a much broader range of information (\eg, videos, web tutorials), in addition to just notebooks.  

Overall, participants conducted similar numbers of code searches with Google Search and \name{} during the study. 
Participants conducted a few more searches with Google Search since they needed to adjust their search keywords several times when the context-based search (in \name) was not available.
For the API recommendation in \name{}, on average participants used it 1.0 times (\sd{0.7}); however, seven out of 14 participants did not use this function. When asked for reasons, they mentioned that it was not needed as they already had an idea of using which methods from their past experience and the received examples from search.   

As the second part of our study task asked participants to try to build a prediction model with the data, we observed 6 out of 14 (43\%) participants successfully employed a machine learning model with Google Search during the study, compared to 10 participants (71\%) with \name{}. This indicates \name{} helped participants build models based on the searched examples.
Moreover, participants used a larger variety of machine learning models with \name{}, including logistic regression, k-means, linear regression, DBSCAN, kNN, random forest, gradient boosting, and decision tree. However, with Google Search, participants only employed k-means, random forest, linear regression, and decision tree.
This may be because the EDA notebooks returned by \name{} have a broader coverage since they are based on participants' customized code; it is more constrained as in Google Search participants searched more similar keywords in the study.

\begin{figure}[!tb]
    \centering
    \includegraphics[width=\linewidth]{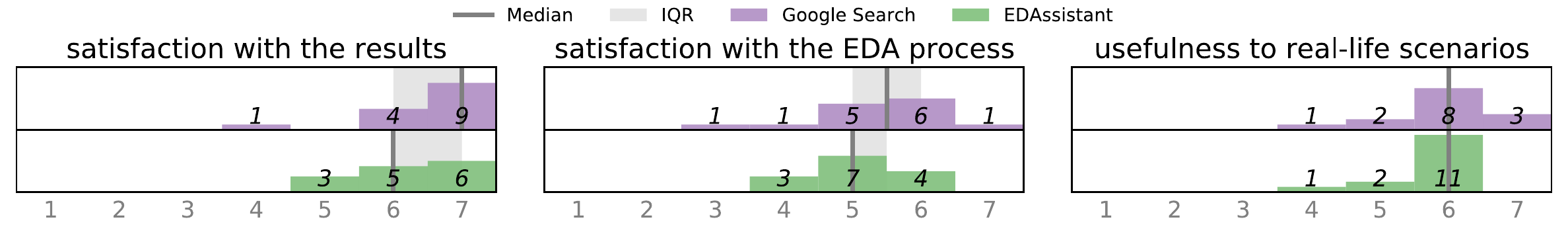}
    \vspace{-7mm}
    \caption{Participants' ratings on the post-study questionnaire on a 1--7 Likert scale (1 - strongly disagree, 7 - strongly agree).} 
    \label{fig:ratings}
\end{figure}

\autoref{fig:ratings} shows participants' ratings on the post-study questionnaire (the higher the better). 
In both conditions, participants were highly satisfied with their EDA results (Google: \md{7}, \iqr{1}; \name: \md{6}, \iqr{1}) and EDA processes (Google: \md{5.5}, \iqr{1}; \name: \md{5}, \iqr{0.5}), where the medians of the ratings on both tools were similar. Google Search was rated slightly higher, and one reason was that it could provide more diverse search results other than just code. 
Moreover, for the usefulness to real-life scenarios, both were rated highly (\md{6}, \iqr{0}), which indicates that participants perceived \name{} as a good alternative compared to Google Search. 


\subsubsection{Qualitative Feedback} \label{sec:feedback}
Overall, participants were positive about their experience with \name{} and appreciated the new ways of conducting EDA and notebook example searches. They thought the interface of \name{} was intuitive and useful. 
\qt{The idea is good and obvious.}-P14.
They also liked the visual encodings and user interactions, which made it easier to understand the searched examples. In the following, we organized our interview results based on the design goals in \autoref{sec:design-goals}.  

\textbf{In-context and in-situ search experience for EDA examples (G1).}
Participants, in general, felt the search experience with \name{} smoother and more effective than using Google Search.
\qt{It was helpful to give me similar code examples.}-P7. \qt{I think the system (\name) was very useful. The relevant libraries are all there.}-P10. 
P4 echoed the same point: \qt{This will definitely help programmers, because we do need examples, not the APIs.}
Further, P12 had the following statement when comparing \name{} with Google Search: \qt{The main advantage of the interface (\name) is that the code it provided was all usable because they are from Kaggle. If I was using Google, I wanted to see the detailed implementation, but normally there might be some non-related texts.} 
Participants also gained experience while using \name{}. For instance, through the searched examples, P7 mentioned \qt{I did not realize I can actually plot two separate charts to solve the first task.}

At the same time, participants pointed out that \name{} did not have the capabilities as Google Search to retrieve a diverse set of results beyond notebooks, such as video tutorials, forum discussions, etc.
\qt{Google will return results with non-code stuff. For instance, some StackOverflow results, which provide code and detailed explanation.}-P9.
\qt{Because Google usually has many different types of results, some of them have interactive explanations which save time for understanding the code.}-P6. 
Participants (P2 and P5) also suggested that it would be more effective to combine the keyword-based search with Google Search and the in-situ search with \name{}, which allows the tool to take more \qt{users' opinion} into consideration.  
P12 explained \qt{The advantage of Google is that I can predict the search results, especially when I know which library to use. If I can have a Google + the interface combination, that would be great.}
P4 also stated that using \name{} more would result in better performance: \qt{If I am familiar with it, it will be extremely helpful.}

\textbf{Visualization and exploration of searched EDA examples (G2).}
Participants especially liked the design of \name{} for supporting the exploration and understanding of searched notebooks. For example, P12 mentioned \qt{Visualization and interface are clean and neat. No useless information.}
P1 appreciated the visual design of the Search Results View, mentioning \qt{I think the colors from the first view can help me filter down for what I am looking for. When I am looking for plotting methods, I tried to read the purple lines first.} 
Further, P2 said: \qt{Amazing, very helpful, especially when you don’t know what the code is doing, the visual encoding can give you some hints.}
Similarly, \qt{Compared to Google, which doesn’t have this color-coded highlighting, this is intuitive.}-P8.

Together with the keywords besides the EDA sequences, participants found that \qt{It can help you locate the answers faster. The tags are pretty good, using colors to encode importance. And the colors from the first view are helpful. They basically try to tell you what kind of operation that line is.}-P10. 
\qt{The words besides can tell you what are the packages being used before clicking it, and hovering over gives you the imports. I think the searched sequences are helpful.}-P7.
However, some participants demanded a higher-level summarization for the EDA sequences with text. 
\qt{The idea is good, but sometimes the keywords were not important enough.}-P9.
\qt{Some keywords, such as ‘df’, do not contain much information.}-P14.
\qt{It might be helpful if you can summarize what the code is doing in natural language, rather than keywords.}-P10.

As for the Notebook Detail View, participants realized its importance and used it often with the Search Results View for exploring the EDA sequences. They thought it was standard and effective for demonstrating the code. They also liked the fold/unfold feature: \qt{It’s good to highlight the specific operation you just clicked from the top view, and use fold/unfold to go over the entire notebook.}-P7.  
Participants suggested that some filtering and searching mechanism could be integrated into both views.
\qt{It will be good if the keywords from the first view can tell me which one contains linear regression, and this will save lots of my time. Otherwise, I need to read the code line by line.}-P1. 
Also, P9 recommended that the same color coding of the Search Results View could be added: \qt{But I want to try to introduce some consistency here. In the first view, you use colors to encode the operation type for each line, is it possible to do the same thing from the second view?}

\textbf{Recommendation of APIs to use next (G3).}
Seven out of 14 participants used the API recommendations in \name{} and they thought this was useful for obtaining the methods to use and checking the API documentation, which was another key advantage participants perceived for our tool.  
\qt{It can quickly go to the documentation page, which is the main advantage. Usually, I click the dark green one first, because it usually gives the most relevant method.}-P7.
\qt{I realized the bottom view later, but the bottom view was really helpful to understand the key methods that are helpful.}-P2.
Moreover, P14 gave specific examples: \qt{I think the bottom view was good. After drawing the first catplot, it recommended me with the matplotlib.plot function for the second chart. After writing some code related to training, it provided me with some validation functions.}
Similar to searching EDA sequences, participants suggested the integration of manual search keywords into the API recommendation. 
\qt{I hope the button could take user text inputs, so that I can have a little bit of control over the results.}-P5.
Further, P11 recommended an interesting feature that \qt{helps auto-fill current working cell} after clicking the tags.

\section{Discussion and Future Directions}
In this section, we discuss some design implications obtained from our study results, limitations of the current \name{} implementation, and future directions to enhance the work.

\subsection{Design Implications}
From our study results, we observe a trade-off between the ``active-style'' search by inputting keywords on Google Search and the ``passive-style'' search in \name{} based on code written by users.  
Participants thought that actively inputting the keywords allowed them to know what to expect.
\qt{Manual search can help us find things match with ideas in our minds.}-P4.
\qt{If I’d like to find things related to the chart, then I will have some expectations in my head.}-P3.
This is essentially helpful when data scientists have a clearer goal of what to do and what to search. This also explains why participants thought Google performed better in the first part of the EDA task in our study, which was more prescribed. 
However, \qt{Manual search is not always better and it only performs well when results contain what we really expect.}-P7. 

In cases that the goals were vague (\eg, in the second part of each study task), participants felt that the experience with \name{}, by just clicking the ``Search for Examples'' button, was more natural and integrated into their workflows with computational notebooks. 
This is also true when users lack adequate domain knowledge or experience on their problems.
Further, as described in our qualitative results, participants hoped to combine both active and passive search styles together in their EDA. 
Interestingly, we did observe that some participants ``hacked'' our tool by creating a new cell containing the keywords they wanted to search with and initiating the search. 
\qt{Later I tried to put some keywords into the current working cell, to see whether I could affect final results.}-P14.
However, this does not fully leverage the advantages of \name{} in understanding the code structures.
Therefore, in the future, it is worth considering how to design such integrated EDA support tools within the Jupyter environment, while balancing the two search styles.  

Moreover, another trade-off lies between the diversity and consistency of search results. 
Several participants (such as P6 and P9) mentioned that Google retrieved a lot of \qt{non-code stuff} such as forum discussions and videos, which could benefit their EDA tasks. 
For example, a few well-curated interactive tutorials or web blogs could be significantly helpful for learning new concepts and APIs. However, this highly depends on the availability of the resources and the specific cases that users encounter.
On the other hand, due to limited data diversity, \name{} right now can only retrieve EDA notebook examples. P8 made a nice analogy that \qt{I think what you are trying to replicate is like what I was doing: when I forgot how to plot chart, I would search among my past projects and find what I did.}
Indeed, in some corporate scenarios, teams tend to work on a set of aligned goals and use many internal APIs of which learning materials are hard to find in the public domain \cite{li2021nbsearch}. 
More consistent and focused search results would be beneficial for onboarding new employees or team members, which may otherwise overwhelm the users.
Also, developers in companies usually generate a lot of code but few well-made tutorial resources due to various constraints on time and money. 
Being able to search for code examples by other senior co-workers in a seamless way would increase new members' productivity under such constraints.
Thus, future research could be conducted to study how the diversity or consistency factor affects data scientists' EDA in different real-world scenarios.

Finally, there exists a trade-off between encouraging creativity and following past practices for novice data scientists using tools like \name. 
As the suggested EDA sequences are mined from Kaggle competitions, this might compel data scientists to follow certain routines of analysis. While EDA is a relatively creative process, these suggestions may further reinforce common previous practices and allow more and more users to follow. 
This is a double-edge sword. On one hand, it helps novice data scientists quickly gain skills from existing knowledge, but on the other hand, it reduces their chances to experiment new ideas and approaches in EDA. 
While assessing the trade-off is out of the scope of this paper, it is a profound research problem that should be studied in-depth in the future.

\subsection{Limitations and Future Work} 

Our tool and study design are not without limitations. 
First, while we trained our models on a reasonable-sized corpus including around 38K notebooks, it is still small compared to the vastly available online resources that other search engines (\eg, Google Search) can leverage. Further, all the notebooks were captured from competitions on Kaggle to ensure their quality, this, however, may introduce bias into the trained models. 
Thus, further collecting larger and more diverse data is necessary to maximize the potentials of effective in-situ search experience that \name{} has brought to data scientists. 
Using more diverse data, we could also support data scientists with better understanding why certain notebooks or APIs are suggested (\ie, explaining the recommendation), which is a promising future direction. 

Moreover, EDA is highly dynamic and case by case, and the notebooks collected in Kaggle competitions are dataset or problem-dependent. Data scientists may work on a different problem and at the same time require suitable examples. 
While the literature \cite{yan2020autosuggest,barel2020automatically} has attempted to learn data-independent patterns from notebooks and the models we used have such abilities, the utility of retrieved example notebooks can still vary in different situations. P9 mentioned that \qt{People can have different next steps because they have different goals.}  
The characteristics of working datasets have not been considered in \name{}'s search and recommendation yet. 
Future approaches for learning embeddings that also represent dataset features can be employed, such as in a similar vein of VizML \cite{hu2019vizml} for generating visualizations from datasets.

Third, as discussed above and in participants' feedback (\autoref{sec:feedback}), the current code search and recommendation in \name{} lack diversity in results and finer manual control over the inputs. 
Computational methods and visual interfaces to support these functions can be developed in the future. 
Also, participants pointed out some other improvements for \name{} to better support the understanding of the searched examples, such as providing a natural language code summary instead of discrete keywords in the Search Results View. There is a need for \name{} to offer more context for the retrieved EDA code, such as relating to forum discussions and video tutorials that are sometimes returned in Google Search.  
Thus, it is interesting to broaden our corpus to include these contents on top of notebooks as well as develop code summarization capability of machine learning models (\eg, \cite{zhang2020retrieval}).

Last, our study still has limitations, since EDA is often open-ended and flexible. The current study design is in a controlled environment with fixed datasets, problems, and task procedures, whereas EDA in the wild can be more diverse. Longer-term deployment studies are needed to thoroughly examine the strengths and weaknesses of \name{}, compared to data scientists' existing practice. 
Also, as discussed above, the proposed approach could be more effective in corporate scenarios where employees deal with similar sets of problems, compared to more general-purpose search cases. 
Future studies need to be conducted to examine this hypothesis.

\section{Conclusion}
We have presented \name{}, an interactive and visual tool that facilitates EDA with in-situ code search, exploration, and recommendation, which is developed as a JupyterLab extension to enable a seamless user experience. 
To develop \name{}, a large corpus of high-quality EDA notebooks was curated from the competitions on Kaggle. 
We then characterized data scientists' behaviors with EDA notebooks based on a qualitative formative study and a quantitative analysis of the corpus. 
Advanced machine learning models were trained and evaluated based on the corpus, resulting in the search and recommendation modules of the tool. 
\name{} also features a novel visualization to support the exploration and understanding of searched EDA examples, as well as a user-friendly interface for accessing the search and recommendation functionalities. 
A user study was conducted to assess the strengths and weaknesses of \name{} in EDA tasks as well as a baseline setup of using the external Google search. 
The results indicate that, while Google Search performed better in search results diversity and input control, participants appreciated the \name{} design and the EDA experience with the tool as well as seemed more successful in building prediction models in EDA tasks. 

\bibliographystyle{ACM-Reference-Format}
\bibliography{references.bib}

\end{document}